\documentstyle[twoside,ijmpc2,graphicx]{article}
\ifx\pdfimage\undefined
\else
\fi
\catcode`\@=11
\long\def\@makefntext#1{
\protect\noindent \hbox to 3.2pt {\hskip-.9pt  
$^{{\eightrm\@thefnmark}}$\hfil}#1\hfill}		

\def\ps@myheadings{\let\@mkboth\@gobbletwo
\def\@oddhead{\hbox{}
\rightmark\hfil\eightrm\thepage}   
\def\@oddfoot{}\def\@evenhead{\eightrm\thepage\hfil
\leftmark\hbox{}}\def\@evenfoot{}
\def\sectionmark##1{}\def\subsectionmark##1{}}

\begin{document}

\runninghead{Gerald H. Ristow} 
     {Tumbling Motion of Elliptical Particles in Viscous Two-Dimensional Flow}

\normalsize\textlineskip
\thispagestyle{empty}
\setcounter{page}{1}

\copyrightheading{Vol. 11, No. 0 (2000) 000--000}

\vspace*{0.88truein}

\fpage{1}
\centerline{\bf TUMBLING MOTION OF ELLIPTICAL PARTICLES}
\vspace*{0.035truein}
\centerline{\bf IN VISCOUS TWO-DIMENSIONAL FLUIDS}
\vspace*{0.37truein}
\centerline{\footnotesize GERALD H. RISTOW}
\vspace*{0.015truein}
\centerline{\footnotesize\it Fachrichtung Theoretische Physik, Universit\"at
des Saarlandes}
\baselineskip=10pt
\centerline{\footnotesize\it Postfach 15 11 50, 66041 Saarbr\"ucken, Germany}
\vspace*{0.225truein}
\publisher{(February 24, 2000)}{(April 20, 2000)}

\vspace*{0.21truein}
\abstracts{The settling dynamics of spherical and elliptical particles in a
           viscous Newtonian fluid are investigated numerically using a finite 
           difference technique. The terminal velocity for spherical particles
           is calculated for different system sizes and the extrapolated value
           for an infinite system size is compared to the Oseen approximation.
           Special attention is given to the settling and tumbling motion of
           elliptical particles where their terminal velocity is compared with
           the one of the surface equivalent spherical particle.}{}{}



\vspace*{1pt}\textlineskip	
\section{Introduction}		
\vspace*{-0.5pt}
\noindent
Computational fluid dynamics is a very fascinating and still rather challenging
field of study. This is especially true when solid particles are immersed in a
viscous fluid leading to a two-phase flow problem which can be dealt with in two
fashions: (a) Through a continuum approach involving both phases simultaneously
or (b) by solving the governing equations separately for each phase and
combining the results appropriately.

In this article, we will use the latter approach and study the sinking and
tumbling motion of spherical and elliptical particles in two-dimensional
spatially bound viscous fluids. The fluid motion is solved on an equally-spaced
finite difference grid and the particle-fluid interaction is taken into account
by (a) treating the particle as an additional moving boundary and (b) by
integrating the stress tensor over the surface of the particle in order to
calculate the force and torque on the particle. The details of the algorithm
are outlined in Sect.~2. The results of our numerical scheme are validated in
Sect.~3 by comparing them with the Oseen approximation in the low Reynolds
number limit. In Sect.~4, the settling and tumbling dynamics of elliptical
particles are studied for different aspect ratios and system sizes. Special
focus is given to the magnitude and the decay of the oscillations in the
tumbling motion of the ellipses. Our results are summarized in Sect.~5.

\section{Numerical Modeling}
\noindent
A finite difference technique is used to study the sinking dynamics of spherical
and elliptical particles in two-dimensional, bounded domains filled with a
viscous fluid.

\ifx\pdfimage\undefined
\else
  \pagebreak
\fi

\subsection{Governing equations}
\noindent
Since the obtained velocities are well-below the speed of sound, the fluid can
be treated as incompressible, i.e.\ having a constant density $\rho_f$. The
fluid velocity $\vec{v}$ and the pressure $p$ are measured in a fixed
laboratory system and the governing equations are the Navier-Stokes equations
which read in dimensionless form

\begin{equation}
  \frac{\partial \vec{v}}{\partial t} + (\vec{v}\cdot\nabla)\vec{v} = 
  -\nabla p + \frac{1}{\mbox{Re}} \nabla^2 \vec{v} + f\ ,
  \label{eq: navier-stokes}
\end{equation}
where $\mbox{Re}$ denoted the Reynolds number and $f$ an external force, e.g.\
gravity $g$.

The only parameter that describes the physical situation is the dimensionless
Reynolds number Re. Given a characteristic velocity $U$, a characteristic
length $D$, e.g.\ the particle diameter, and the viscosity of the fluid $\nu :=
\eta/\rho_f$ its definition is
\begin{equation}
   \mbox{Re} := \frac{U \, D}{\nu} \ .
   \label{eq: re}
\end{equation}

For an incompressible fluid the continuity equation
\begin{equation}
  \frac{\partial \rho_f}{\partial t} + \nabla\cdot (\rho_f \vec{v}) = 0
  \label{eq: continuity2}
\end{equation}
reduces to
\begin{equation}
  \nabla \cdot \vec{v} = 0\ .
  \label{eq: continuity}
\end{equation}

We study the motion of rigid particles in fluids by introducing the particles
as additional, moving boundaries.\cite{ristow96} The force acting on the
particle by the fluid is calculated via integration of the stress tensor
$\sigma$ along the particle's circumference,\cite{davis95}
\begin{equation}
  \vec{F}_s = \int_{\mbox{\small circumference}} \sigma\cdot\vec{S}\, dA\ ,
  \label{eq: stress1}
\end{equation}
where $\vec{S}$ stands for the outward-pointing surface normal along the
circumference and
\begin{equation}
  \sigma = -p\, I + \eta\left(\nabla\vec{v} + (\nabla\vec{v})^t\right)
  \label{eq: stress2}
\end{equation}
using a matrix notation with $I$ being the unity matrix and $\eta$ the fluid
viscosity. Instead of integrating Eq.~(\ref{eq: stress1}) directly along the
particle circumference as proposed by our algorithm, an analytic expansion for
the fluid field can also be used which is not as straight forward and more
demanding on the computer resources.\cite{kalthoff97}

When Eqs.~(\ref{eq: navier-stokes}) are discretized in time first, $t = t_0 + n
\Delta t$, the general form reads~\cite{peyret85}
\begin{eqnarray}
   \frac{\vec{v}^{n+1} - \vec{v}^n}{\Delta t} &+& \theta \left[ (\vec{v}^{n+1}
   \nabla) \vec{v}^{n+1} - \frac{1}{\mbox{Re}}\nabla^2\vec{v}^{n+1} -
   \vec{f}^{n+1} \right] \nonumber \\
   &+& (1-\theta)  \left[ (\vec{v}^n \nabla) \vec{v}^n -
   \frac{1}{\mbox{Re}}\nabla^2\vec{v}^n - \vec{f}^n \right] +
   \nabla p^{n+\tau} = 0 \ ,
   \label{eq: ns_dis}
\end{eqnarray}
where the parameter $\theta$ fulfills $0 < \theta < 1$. For $\theta = 0$ we end
up with an {\em explicit} scheme and for $\theta = 1$ we get a {\em fully
implicit} scheme. A very efficient method results for $\theta=\frac{1}{2}$
called the {\em Crank-Nicolson Scheme} which we will use to investigate the
dynamics of the fluid. The parameter $\tau$ is set to the value
that will give the lowest truncation error for the desired algorithm. 

One writes the Navier-Stokes equations and the continuity equation in the new
variables $v_x^{n+1}$, $v_y^{n+1}$ and $p^{n+\tau}$ and ends up with a
nonlinear algebraic system for the unknowns
\begin{eqnarray}
   L_x(v_x^{n+1},v_y^{n+1},p^{n+\tau}) &=& 0 \nonumber \\
   L_y(v_x^{n+1},v_y^{n+1},p^{n+\tau}) &=& 0 \label{eq: num_1} \\
   D(v_x^{n+1},v_y^{n+1}) &=& 0 \ . \nonumber
\end{eqnarray}

This system is solved by an iterative procedure where $m$ denotes the iteration
index
\begin{eqnarray}
   v_x^{m+1} - v_x^m + \kappa L_x(v_x^{n+1},v_y^{n+1},p^{n+\tau}) &=& 0
   \nonumber \\
   v_y^{m+1} - v_y^m + \kappa L_y(v_x^{n+1},v_y^{n+1},p^{n+\tau}) &=& 0
   \label{eq: num_2} \\
   p^{m+1} - p^m + \lambda D(v_x^{n+1},v_y^{n+1}) &=& 0 \ , \nonumber
\end{eqnarray}
and the parameters $\kappa$ and $\lambda$ are chosen in such a way that
convergence of this scheme is guaranteed.

From a linear stability analysis one obtains as necessary condition for
convergence of the procedure given by Eqs.~(\ref{eq: num_2}) the condition
\begin{equation}
   \frac{\kappa}{\Delta x^2} \left( \frac{4\theta}{\mbox{Re}} + \frac{\Delta
   x^2}{2 \Delta t} + 2\lambda \right) \le 1 \ , \ \kappa > 0, \lambda > 0 \ .
   \label{eq: num_3}
\end{equation}
It can be obtained in the same way as for the artificial compressibility
method.\cite{peyret85}

For given values of $\Delta t$, $\Delta x$ and $\kappa$ Eq.~(\ref{eq: num_3})
gives an upper bound for the iteration parameter $\lambda$

\begin{equation}
   \lambda < 
   \lambda{\mbox{\scriptsize max}} := \frac{\Delta x^2}{2\kappa} - \left(
   \frac{2\theta}{\mbox{Re}} + \frac{\Delta x^2}{4\Delta t} \right) \ .
   \label{eq: num_4}
\end{equation}

For each iteration it is crucial for a fast convergence of the iterative
algorithm that optimal values for the parameters $\kappa$ and $\lambda$ are
used. Peyret and Taylor~\cite{peyret85} pointed out that the best value of
$\lambda$ is close but not equal to $\lambda{\mbox{\scriptsize max}}$ and the
simulations for our system confirmed this. But the rule of thumb that
$\kappa_{\mbox{\scriptsize opt}}$ should be of the order of $\Delta x^2$ was
only true in our case when the particle was fixed in space. When the particle
was allowed to move according to the calculated force no easy scaling law could
be found since the value depends on $\Delta x$, $\Delta t$ and
$\eta$. In this case, the optimum values for the parameters were determined by a
few test runs.\cite{ristow96} 

\begin{figure}[tp]
  \ifx\pdfimage\undefined
  \begin{center}
    \includegraphics[width=0.75\textwidth]{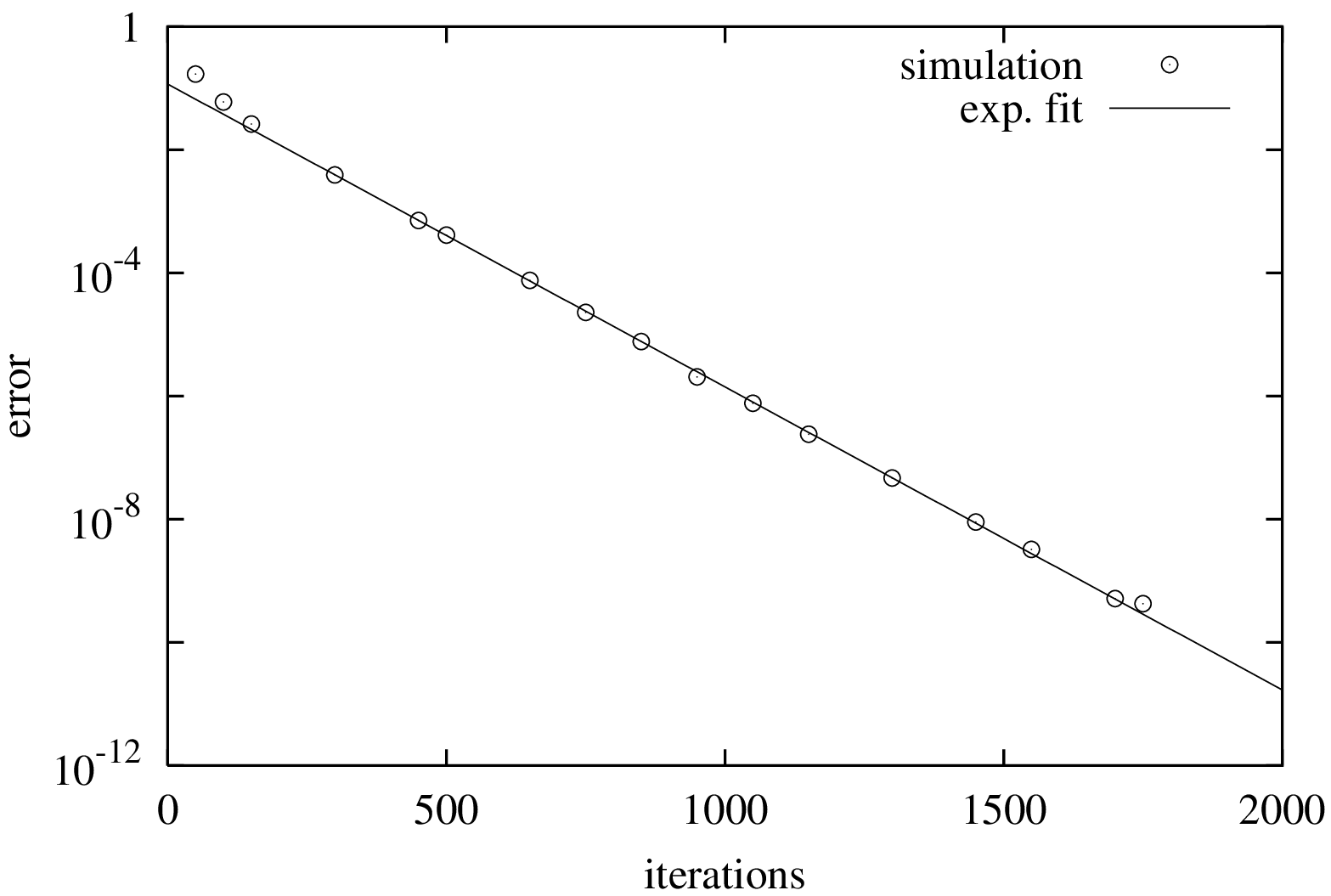}
  \end{center}\vspace{1ex}
  \else
  \begin{center}
    \includegraphics[width=0.75\textwidth]{multi_1.pdf}
  \end{center}\vspace{1ex}
  \fi
  \fcaption{Accuracy in the calculation of the velocity field as function
            of the number of iterations, measured via the divergence of the 
            velocity. The solid line corresponds to an exponential fit with 
            an exponent of $-1$.}
  \label{fig: convergence}
\end{figure}

The convergence rate of our algorithm is illustrated by studying the sinking
dynamics of a cylinder with a radius of 1.5\,cm under gravity using
$g=981$\,cm/s$^2$ released in a two-dimensional box with dimensions 10x20\,cm.
The particle is 10\% denser than the fluid and the fluid viscosity was set to
100 times the value of water. First, we show in Fig.~\ref{fig: convergence} the
number of iterations needed by our algorithm to calculate the velocity field up
to a desired accuracy for a grid size of 64x128 equally-spaced points. The
solid line corresponds to an exponential fit with an exponent of $-1$ and one
reads off that the accuracy increases by one order of magnitude every 200
iterations.

\begin{figure}[t]
  \ifx\pdfimage\undefined
  \begin{center}
    \includegraphics[width=0.75\textwidth]{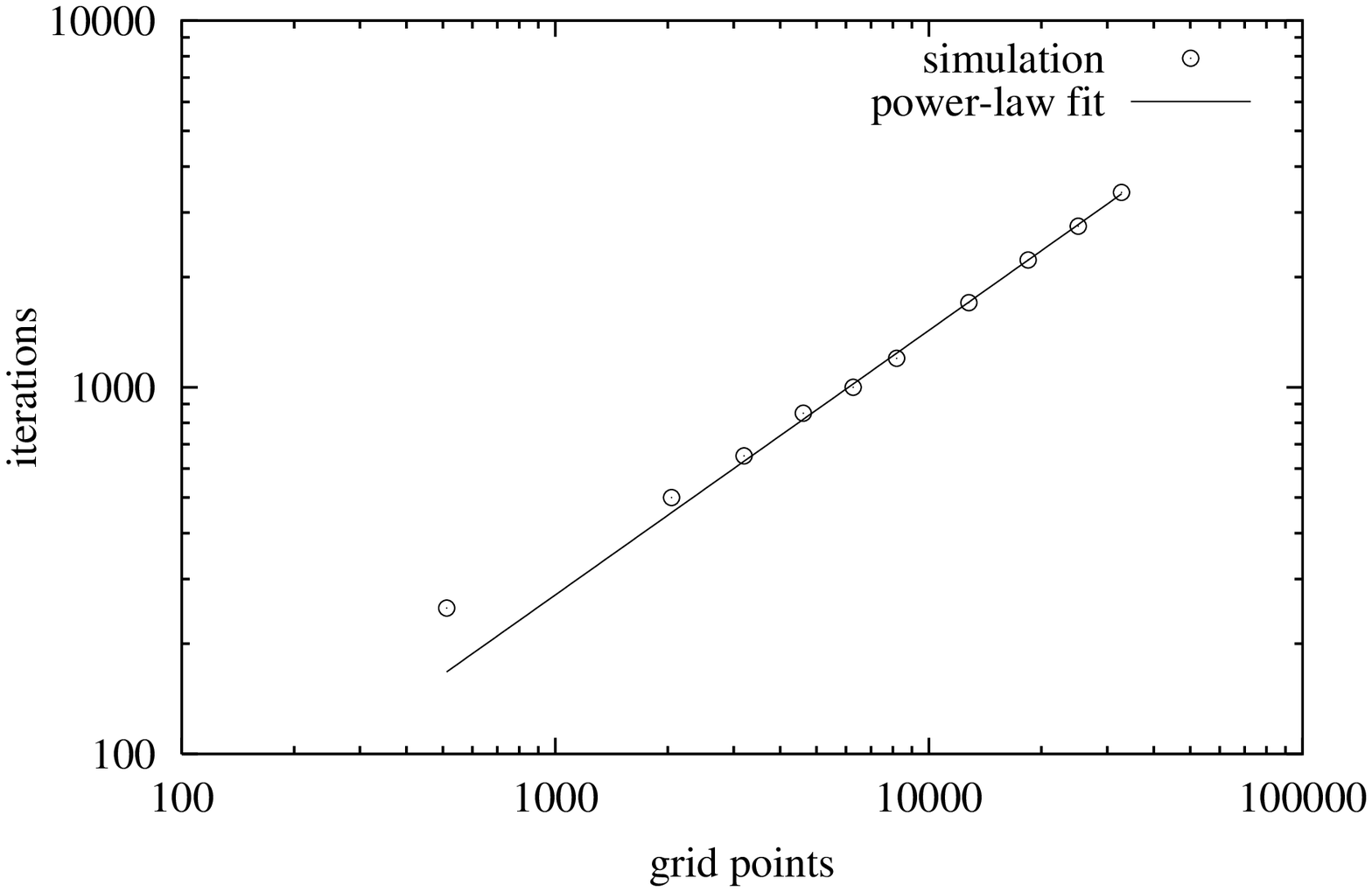}
  \end{center}\vspace{1ex}
  \else
  \begin{center}
    \includegraphics[width=0.75\textwidth]{multi_2.pdf}
  \end{center}\vspace{1ex}
  \fi
  \fcaption{Number of iterations for a desired accuracy as function of the
            number of grid points. The solid line corresponds to a power-law 
            fit with an exponent of $0.72$.}
  \label{fig: grid_dependence}
\end{figure}

For iterative algorithms, the number of iterations to obtain a desired accuracy
usually increases with the number of unknowns, here grid points. This
dependence is shown in Fig.~\ref{fig: grid_dependence} for our algorithm using
the same system parameters as described above and an accuracy of $10^{-7}$. The
number of iterations increases with the number of grid points, as expected, and
the data  points corresponding to more than 2000 grid points could be
well-described by a power-law fit with an exponent of $0.72$. Please note that
the total runtime of the algorithm will exhibit a scaling relation with an
exponent of $1.72$ since the time required for one iteration usually increases
linearly with the number of unknowns.

\subsection{Spherical particles}
\begin{figure}[t]
  \ifx\pdfimage\undefined
  \begin{center}
    \includegraphics[width=0.49\textwidth]{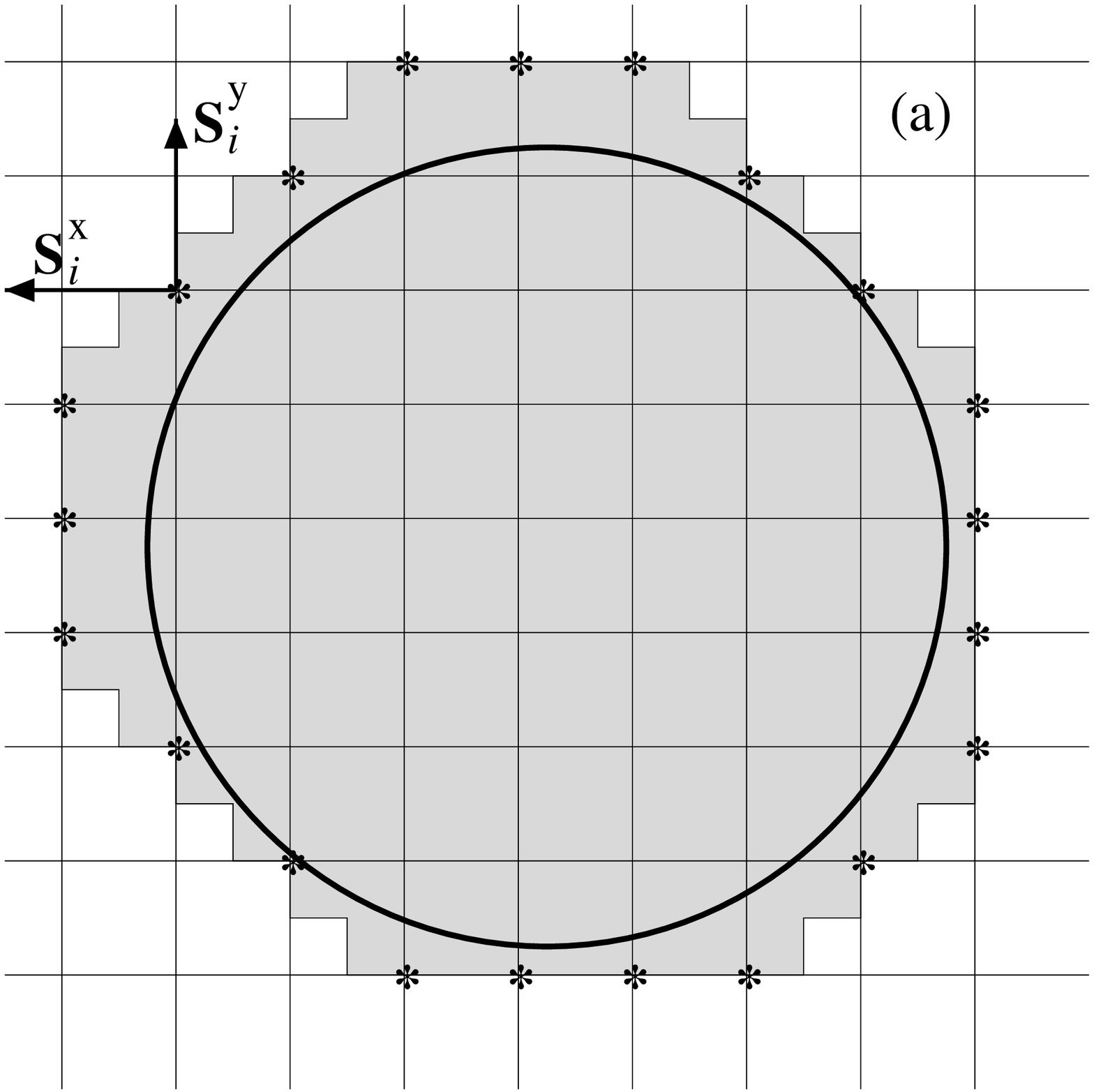}
    \hfill
    \includegraphics[width=0.49\textwidth]{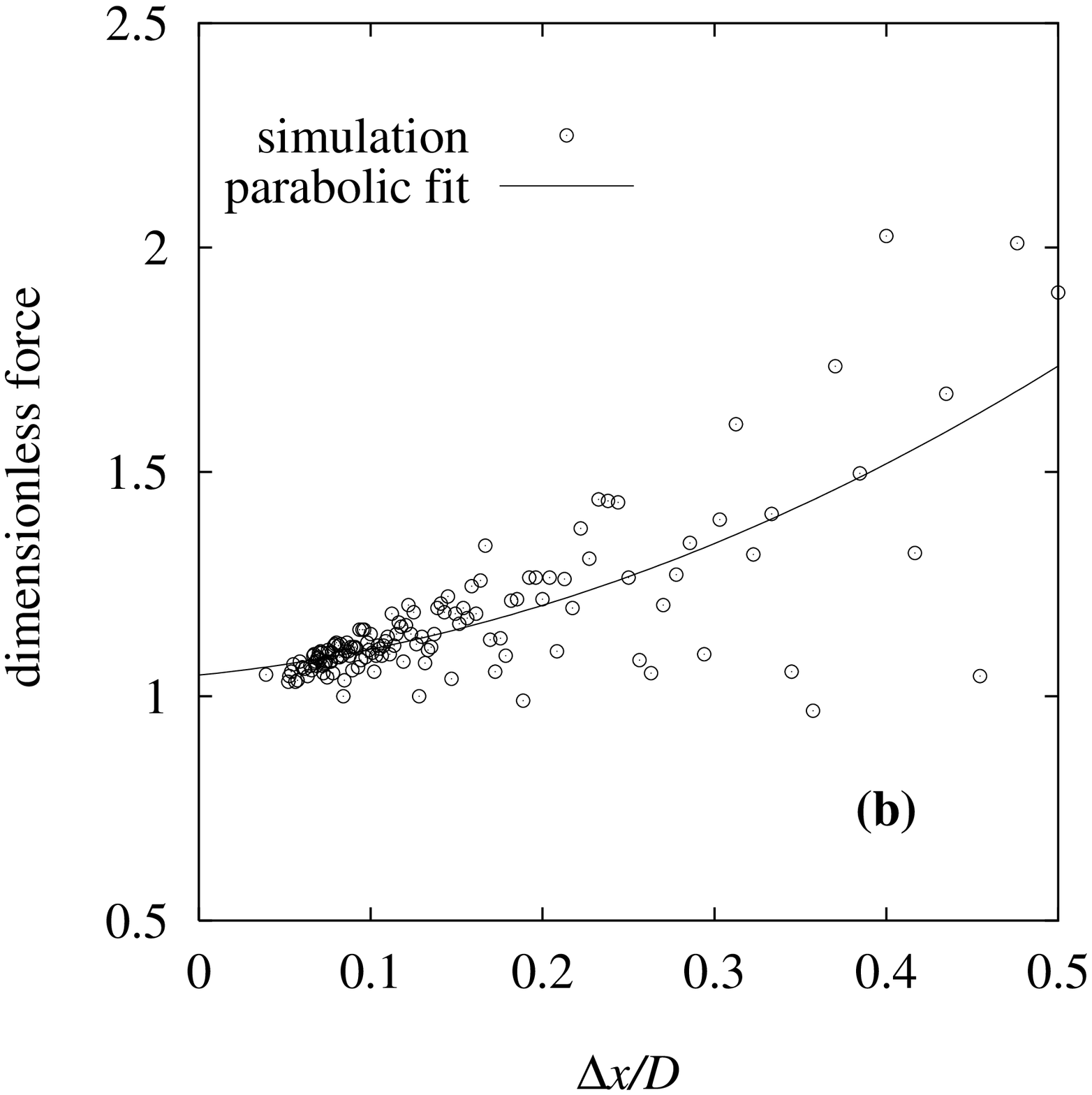}
  \end{center}\vspace{1ex}
  \else
  \begin{center}
    \includegraphics[width=0.49\textwidth]{pressure1.pdf}
    \hfill
    \includegraphics[width=0.49\textwidth]{finite2.pdf}
  \end{center}\vspace{1ex}
  \fi
  \fcaption{Straight-forward approach to calculate the force on a particle: (a)
  Pressure points, denoted by $\star$, used in the calculation and (b) force on
  the cylinder as function of the dimensionless grid spacing. The gray-shaded 
  area illustrates the grid-dependence of the {\em real}\/ particle size and the 
  two decoupled surface normals are also shown in the upper left corner.}
  \label{fig: pressure_old}
\end{figure}

\noindent
As noted earlier, the motion of the rigid particles is treated as an
additional, moving boundary. A straight-forward approach to obtain the pressure
term on the staggered grid is sketched in Fig.~\ref{fig: pressure_old}a, where
the shape of the particle is approximated by boxes (or squares if the grid
spacing is the same in both coordinate directions) and the force on the
particle is calculated by summing over the outer edges of the boxes along the
Cartesian coordinate directions. The two directions decouple when one first
sums along the horizontal and later along the vertical lines, indicated by the
two arrows in the upper left corner. Since the fluid pressure is not
well-defined inside a rigid particle, only points {\em outside}\/ the particle
are used for calculating the pressure. This makes the particle appear {\em
larger}\/ than it really is, shown as gray-shaded area in Fig.~\ref{fig:
pressure_old}a. Consequently, the calculated force is usually over-estimated
but approaches the theoretical value as the grid spacing decreases. This is
shown in Fig.~\ref{fig: pressure_old}b for a cylinder at rest with a diameter
of 2\,cm in a box of 20x40\,cm. The theoretical value for the buoyancy force
per unit length is $F_s=-3.1$\,g/s$^2$ which is reached within a 5\% error for
a spacing of 0.1 which corresponds to more than 100 grid points in the
x-direction and roughly 10 points across the particle diameter. Also shown as
solid line in Fig.~\ref{fig: pressure_old}b is a parabolic fit to the data
points which illustrates the approach to the theoretical value with decreasing
grid spacing. The velocity gradients needed in Eq.~(\ref{eq: stress2}) to
complete the calculation of the force on the particle are calculated by
choosing the appropriated grid points from the staggered MAC grid around the
grid points that entered the pressure calculation.\cite{peyret85} A somewhat
similar approach was used in a multi-particle simulation to study the dynamics
of suspensions.\cite{wachmann98}

The above-described algorithm for calculating the force exerted by the fluid on
the particle has two drawbacks: (i) Large fluctuations are obtained when the
grid spacing is changed, see Fig.~\ref{fig: pressure_old}b, and (ii) the
corresponding particle size, shown in gray in Fig.~\ref{fig: pressure_old}a,
does not match the physical size of the particle but it is mostly
over-estimated by the numerical algorithm.

\begin{figure}[t]
  \ifx\pdfimage\undefined
  \begin{center}
    \includegraphics[width=0.49\textwidth]{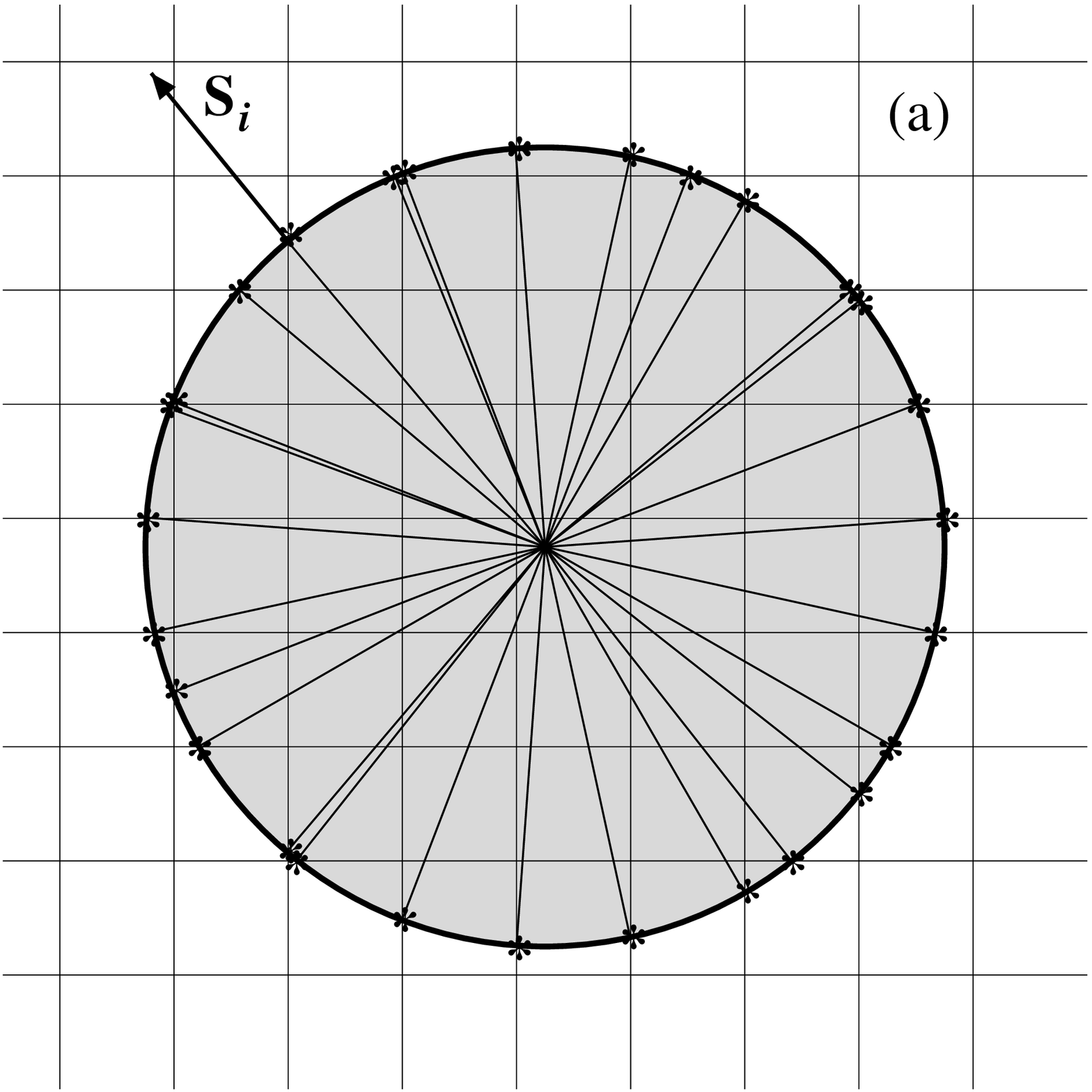}
    \hfill
    \includegraphics[width=0.49\textwidth]{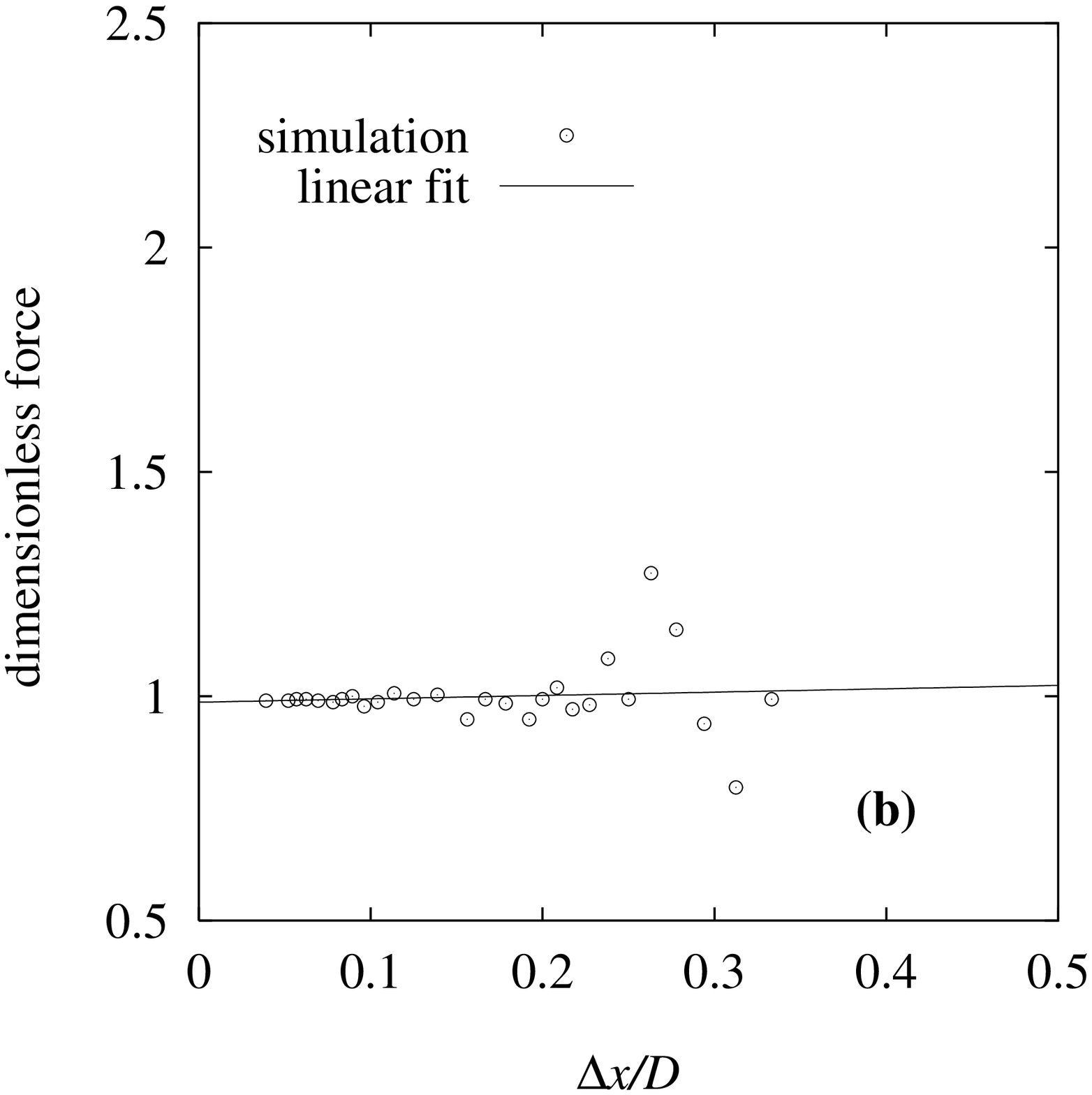}
  \end{center}\vspace{1ex}
  \else
  \begin{center}
    \includegraphics[width=0.49\textwidth]{pressure3.pdf}
    \hfill
    \includegraphics[width=0.49\textwidth]{finite3.pdf}
  \end{center}\vspace{1ex}
  \fi
  \fcaption{Improved approach to calculate the force on a particle: (a)
  Pressure points, denoted by $\star$, used in the calculation and (b) force on
  the cylinder as function of the dimensionless grid spacing. In contrast to 
  Fig.~\ref{fig: pressure_old}a, now the {\em exact}\/ particle size is used for 
  all grid spacings and the surface normal usually has components along both 
  spatial directions, shown in the upper left corner.}
  \label{fig: pressure_new}
\end{figure}

To overcome these problems, a two-step procedure was used. Realizing that the
only contribution to the stress tensor in Eq.~(\ref{eq: stress2}) for a
particle at rest comes from  the pressure $p$, correcting these contributions
seemed favorable. First, the pressure is linearly  extrapolated to the {\em
real}\/ particle surface, indicated in Fig.~\ref{fig: pressure_new}a by
$\star$, thus making the errors in the particle size much smaller. Second,
instead of working with rectangular blocks which decouple the x- and
y-components of the force contributions, the force calculation at each sample
point uses the {\em actual}\/ surface normal, which is indicated by $\mbox{\bf
S}_i$ in Fig.~\ref{fig: pressure_new}a. All surface points are kept in a list
ordered by their arc length (angle). The length of the surface element at each
sample point then uses the averaged arc length to their left and right neighbor
points. This procedure {\em always}\/ leads to a particle size which matches
the physical size exactly in the case when a spherical cylinder is used. The
force fluctuations do no longer show a systematic dependence on the grid
spacing, which is shown in Fig.~\ref{fig: pressure_new}b for the same
parameters as were used to generate Fig.~\ref{fig: pressure_old}b. Also note
that the magnitude of the fluctuations is far less for the improved force
calculation, allowing for a twice as large grid spacing in order to obtain the
same accuracy as before. The extrapolated force for an infinitely dense grid
also gives a more accurate value than the old procedure outlines above.

\subsection{Elliptical particles}
\begin{figure}[t]
  \ifx\pdfimage\undefined
  \parbox[t]{0.495\textwidth}{
    \includegraphics[width=0.49\textwidth]{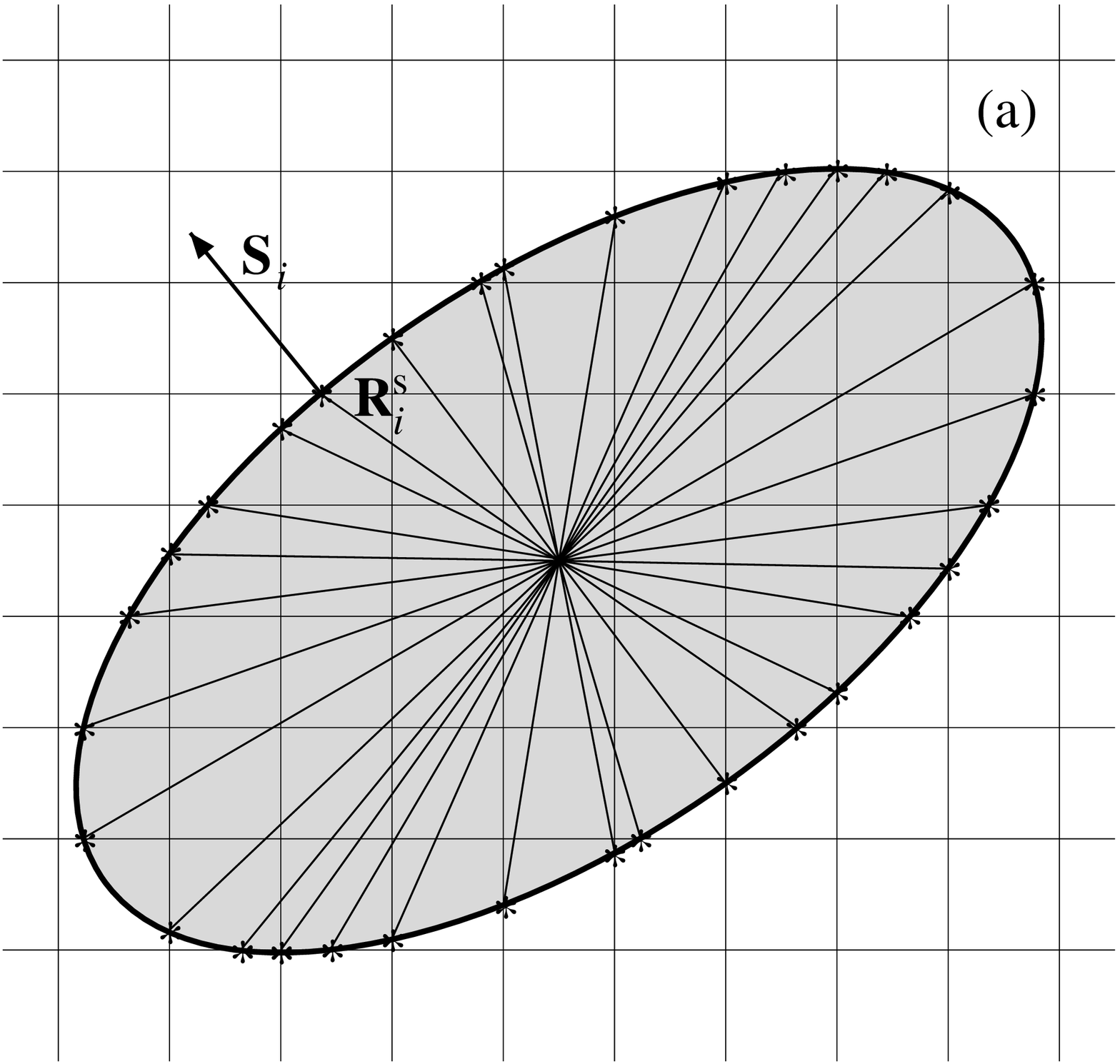}}
    \hfill
  \parbox[t]{0.495\textwidth}{
    \includegraphics[width=0.49\textwidth]{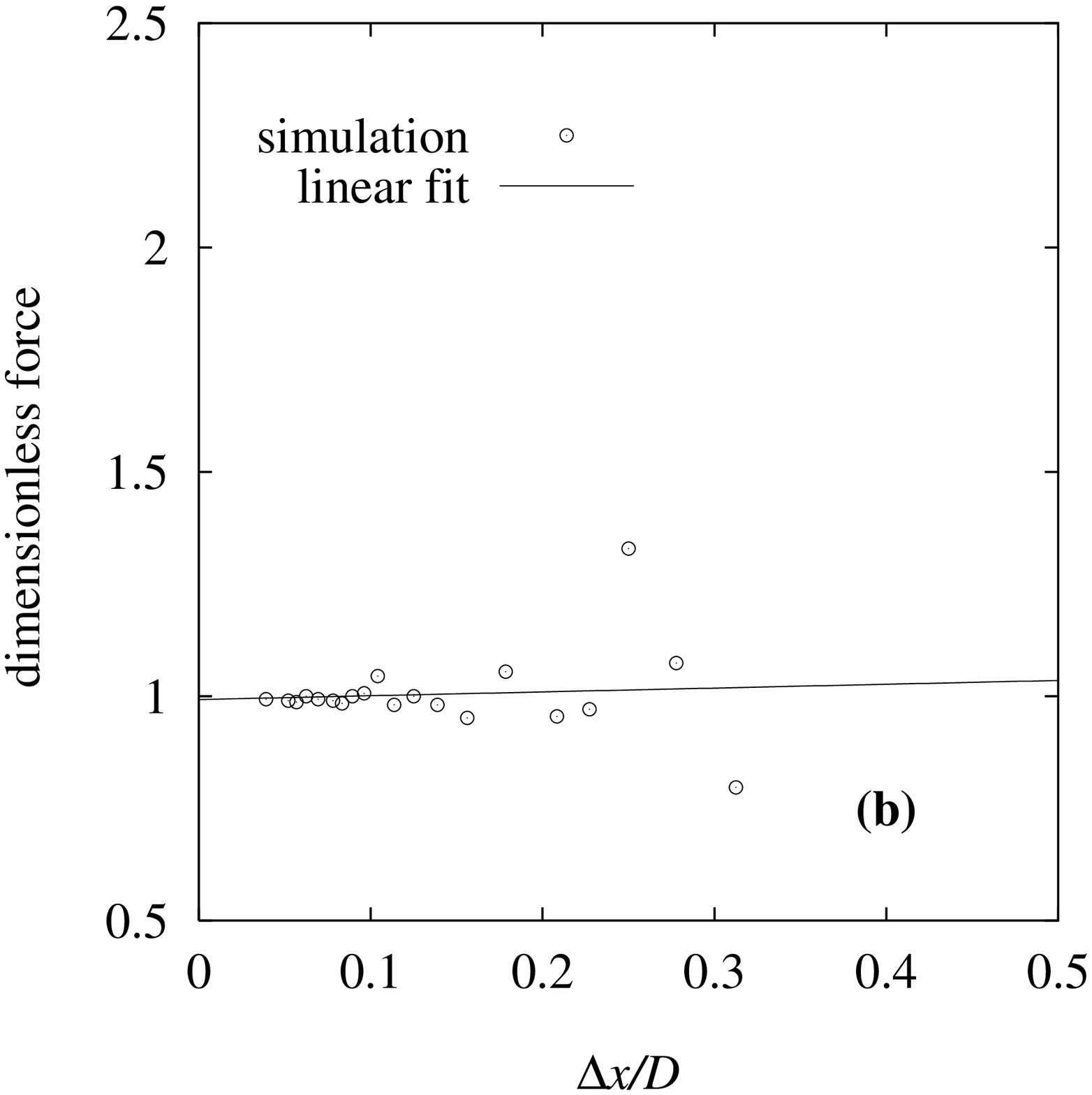} }
  \else
  \parbox[t]{0.495\textwidth}{
    \includegraphics[width=0.49\textwidth]{pressure4.pdf}}
    \hfill
  \parbox[t]{0.495\textwidth}{
    \includegraphics[width=0.49\textwidth]{finite4.pdf} }
  \fi
  \fcaption{Calculating the force on an elliptical particle: (a) Pressure 
  points, denoted by $\star$, used in the calculation and (b) force on
  the ellipse as function of the dimensionless grid spacing.}
  \label{fig: pressure_ell}
\end{figure}

\noindent
As long as spherical cylinders move close to the centerline of the container,
particle rotations can be neglected. This is no longer true if the particle
motion takes place close to one of the side boundaries or if deviations from the
spherical particle shape are considered, e.g.\ elliptical shapes as sketched in
Fig.~\ref{fig: pressure_ell}a. The torque on a particle must then be calculated
and is given by,
\begin{equation}
  \vec{\tau}_s = \int_{\mbox{\small circumference}} \vec{R}^s \times
  (\sigma\cdot\vec{S})\, dA\ ,
  \label{eq: torque}
\end{equation}
where $\vec{R}^s$ denotes the vector from the center of mass of the particle to
the surface element under consideration.

Even though the force (and torque) on an elliptical particle can be calculated
in a similar fashion as described in the preceding section for spherical
particles, two additional problems arise. First, no simple analytic expression
for the arc length between two arbitrary surface points is known and second,
the surface normal ($\vec{S}$) has to be calculated separately for each
sampling point since it is not parallel to the vector from the center of mass to
the surface element ($\vec{R}^s$), as shown in Fig.~\ref{fig: pressure_ell}a in
the upper left corner.

The latter point was addressed by calculating the surface normal analytically
in the reference frame of the ellipse and by rotating it back to the laboratory
frame. For the calculation of the arc length, a simple linear interpolation to
the two neighboring points was used. Since ellipses are concave objects, this
approach will always slightly {\em underestimate}\/ the surface area of the
ellipse, and we introduced an intermediate point in each direction in order to
reduce this error. The accuracy of this approach is demonstrated in
Fig.~\ref{fig: pressure_ell}b, where we plot the force on an ellipse having an
aspect ratio of 2.25 as function of the dimensionless grid spacing. The ellipse
has the same surface area as the circle used to calculate the points shown in
Fig.~\ref{fig: pressure_new}b above, thus showing the same initial vertical
force contribution due to gravity. The accuracy of the two results is
comparable and is far better than the original algorithm presented in
Fig.~\ref{fig: pressure_old}. The extrapolated force for an infinitely dense
grid is hardly distinguishable from the theoretical value. Thus the presented
force calculation for an elliptical particle can be regarded as being as
accurate as for a spherical particle.

\ifx\pdfimage\undefined
  \pagebreak
\fi

\section{Program Validation}
\noindent
In the preceding paragraphs, we showed that the improved algorithm gives
accurate results for the force exerted on a spherical or elliptical cylinder at
rest, see Figs.~\ref{fig: pressure_new} and \ref{fig: pressure_ell},
respectively. We now turn our attention to a moving particle and demonstrate
that the terminal velocity of a cylinder in a viscous fluid as calculated by
the above described algorithm agrees with the theoretical value in the low
Reynolds number limit. In this limit, the Oseen approximation leads to an
analytic expression for the viscous drag force per unit length of the
cylinder.\cite{lamb97,guyon97} This can be used to derive an {\em implicit}\/
expression for the terminal velocity of a cylinder, $V_c$, which reads
\begin{equation}
  \frac{V_c}{\ln(4\nu/(R V_c)) + 0.5} = \frac{\rho_c - \rho_f}{4\eta}g R^2\ .
  \label{eq: oseen}
\end{equation}
Here $\rho_c$ stands for the density of the cylinder, $R$ denotes its radius
and $\nu \equiv \frac{\eta}{\rho_f}$. No analytic expression exists for
arbitrary Reynolds numbers. After the particle is released from rest, the
calculated force and torque exerted by the fluid on the particle is used to
calculate the particle position, orientation, velocity and angular velocity via
the Verlet algorithm. In the iterations for the fluid field for the next time
step, the particle is treated as an additional moving boundary by giving all
velocity points covered by the particle a value which corresponds to a rigid
particle.

\begin{figure}[t]
  \ifx\pdfimage\undefined
  \parbox[t]{0.495\textwidth}{
    \includegraphics[width=0.49\textwidth]{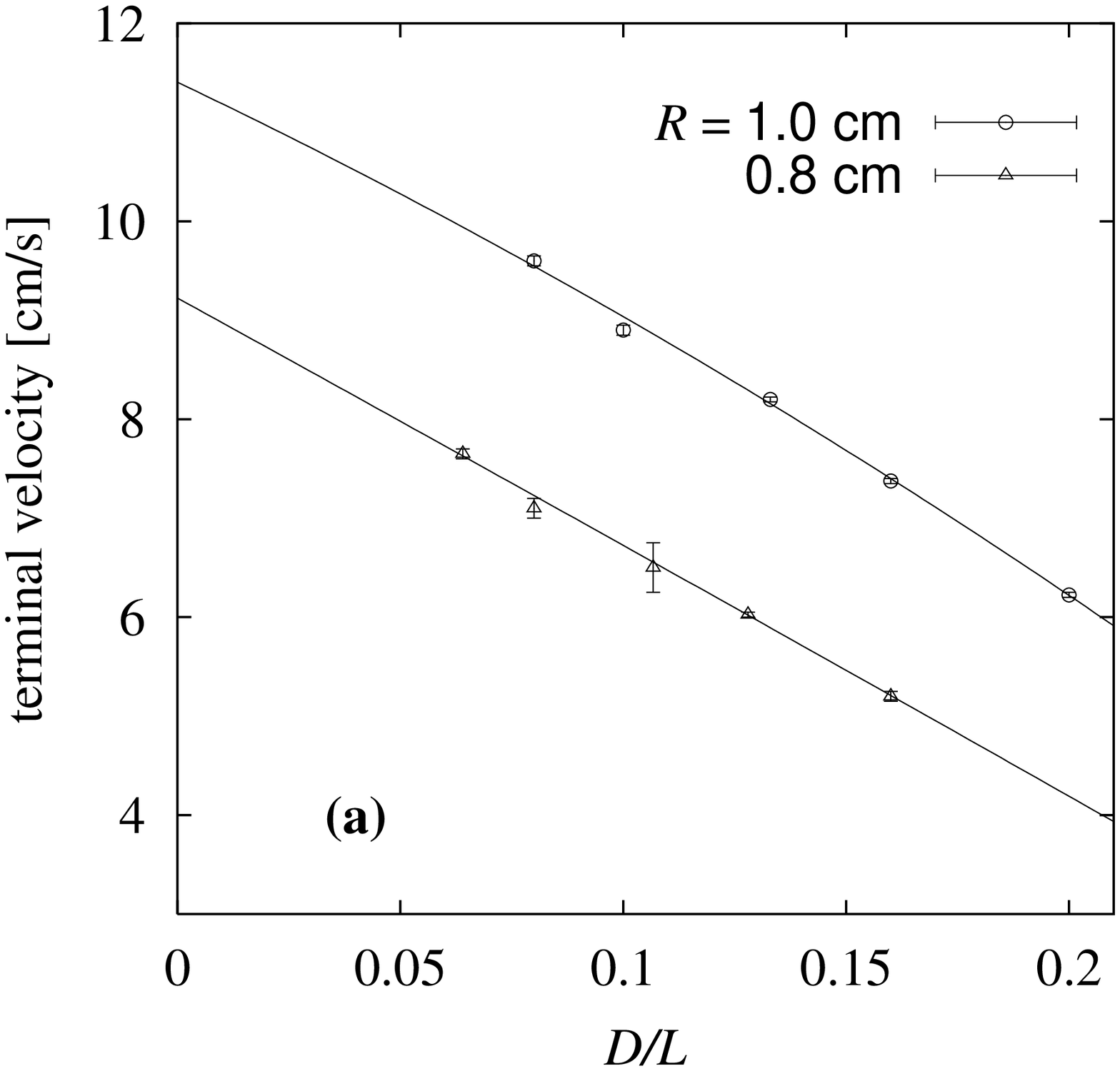}}
    \hfill
  \parbox[t]{0.495\textwidth}{
    \includegraphics[width=0.49\textwidth]{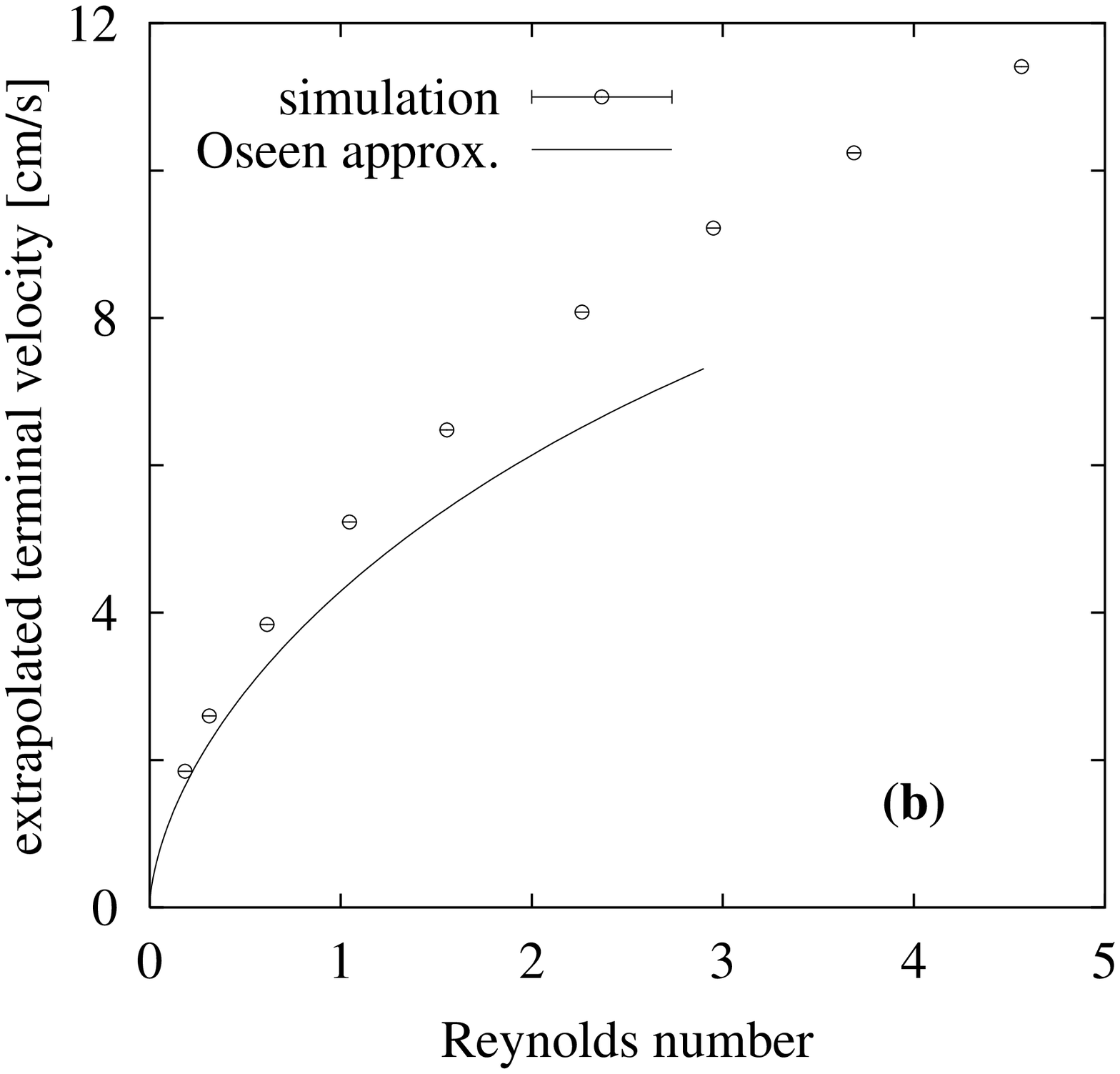} }
  \else
  \parbox[t]{0.495\textwidth}{
    \includegraphics[width=0.49\textwidth]{sink_eta5_inf1.pdf}}
    \hfill
  \parbox[t]{0.495\textwidth}{
    \includegraphics[width=0.49\textwidth]{sink_eta5_inf2.pdf} }
  \fi
  \fcaption{Terminal velocity of cylinders: (a) As function of the diameter to
            system size ratio $D/L$ where the solid lines are quadratic fits to
            the data points and (b) extrapolated value as function of the
            Reynolds number. Also shown as solid line is the Oseen 
            approximation.}
  \label{fig: validate}
\end{figure}

The approximation from Eq.~(\ref{eq: oseen}) is
only valid for an infinite system and we have to extrapolate the numerically
obtained terminal velocities of the cylinder correspondingly.\cite{ristow97}
The procedure we used is illustrated in Fig.~\ref{fig: validate}a for a fluid
viscosity of $\eta=5$\,g/(cm s) and the two cylinder radii 1.0 and 0.8\,cm,
respectively. The density contrast was chosen again as 10\%, see Sect.~2.1. The
terminal velocity is plotted as a function of the dimensionless parameter $D/L$
where $D\equiv 2R$ stands for the cylinder diameter and $L$ for the system
width. A quadratic fit is used to obtain the extrapolated terminal velocities
for an infinite system, i.e.\ in our case the value at the intersection with
the ordinate.

The terminal velocities obtained in this fashion are plotted in Fig.~\ref{fig:
validate}b as a function of the Reynolds number Re. The error bars stem from
the fitting algorithm. Also shown as solid line is the Oseen approximation,
Eq.~(\ref{eq: oseen}). Since the approximation is only valid for low Reynolds
numbers, it is not surprising that the agreement between the numerical results
and the analytic expression becomes better and better the smaller the Reynolds
number gets. However, the smaller the particle becomes the more grid points are
needed to avoid unphysical oscillations in the particle velocity. This only
made calculations down to a cylinder radius of 0.25\,cm feasible due to the
limitations in computer memory and CPU time. Please note that for $R=0.25$\,cm
the corresponding Reynolds number is still moderate (Re$\approx 0.14$) thus
making a value difference of less than 24\% in Fig.~\ref{fig: validate}b for
the left-most data point ($R=0.25$\,cm) more than acceptable.

\section{Motion of Elliptical Particles}
\noindent
When elliptical particles are considered, the particle orientation is important
and particle rotation has to be taken into account as well. In the creeping flow
regime, the particle orientation will in principle not change during the
settling process but the settling velocity depends significantly on the
orientation of the major axis with respect to the container bottom (horizontal
direction).\cite{clift78}

If the {\em major}\/ axis is parallel to gravity, the resistance will be lowest
and the terminal velocity highest. However, such a particle orientation seems to
be unstable for moderate Reynolds numbers where the ellipse prefers to sink with
the lowest possible velocity, thus aligning the {\em minor}\/ axis parallel to
gravity.\cite{cox65,huang94}

\subsection{Terminal Velocity}
\begin{figure}[t]
  \ifx\pdfimage\undefined
  \parbox[t]{0.495\textwidth}{
    \includegraphics[width=0.49\textwidth]{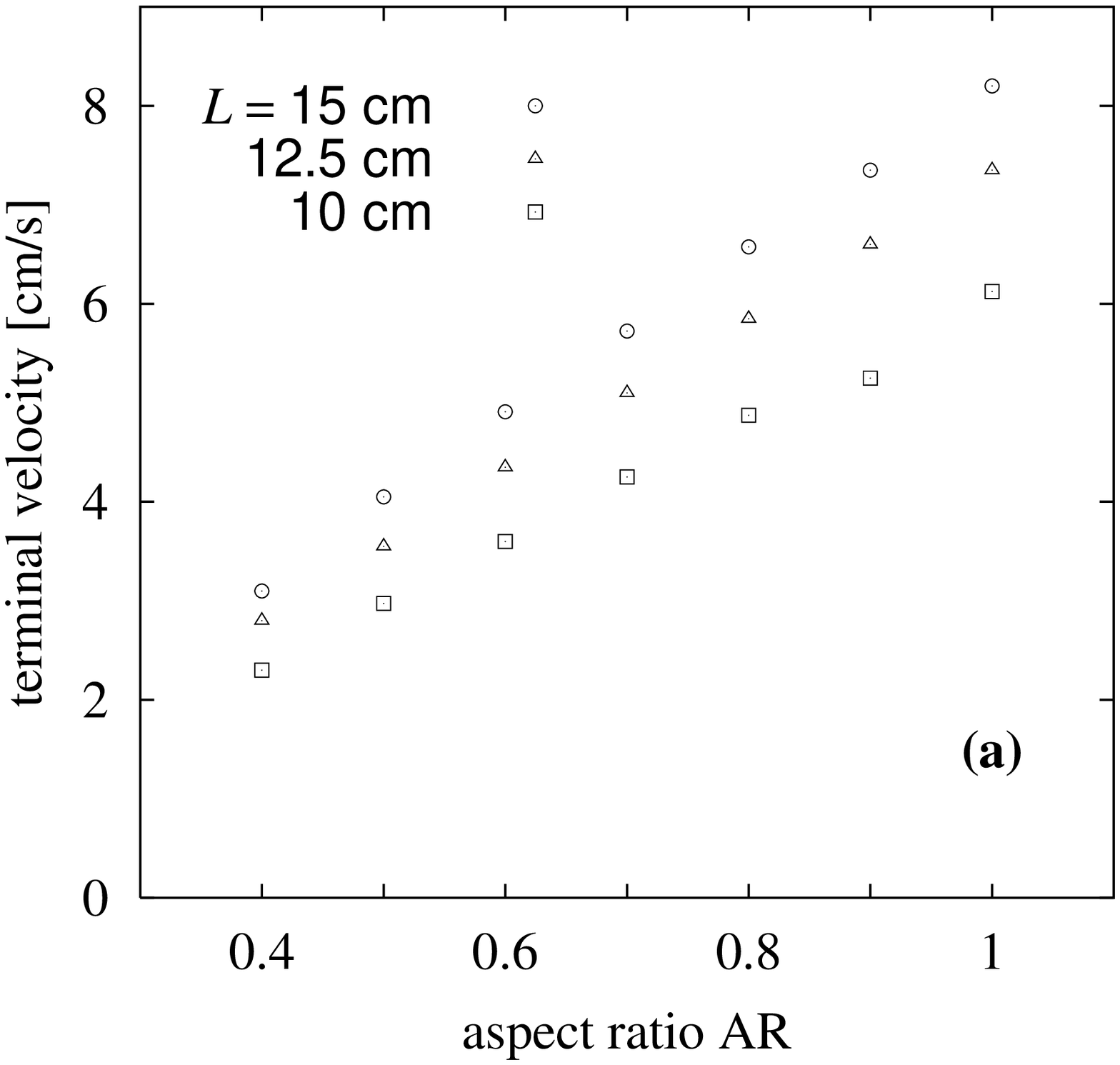}}
    \hfill
  \parbox[t]{0.495\textwidth}{
    \includegraphics[width=0.49\textwidth]{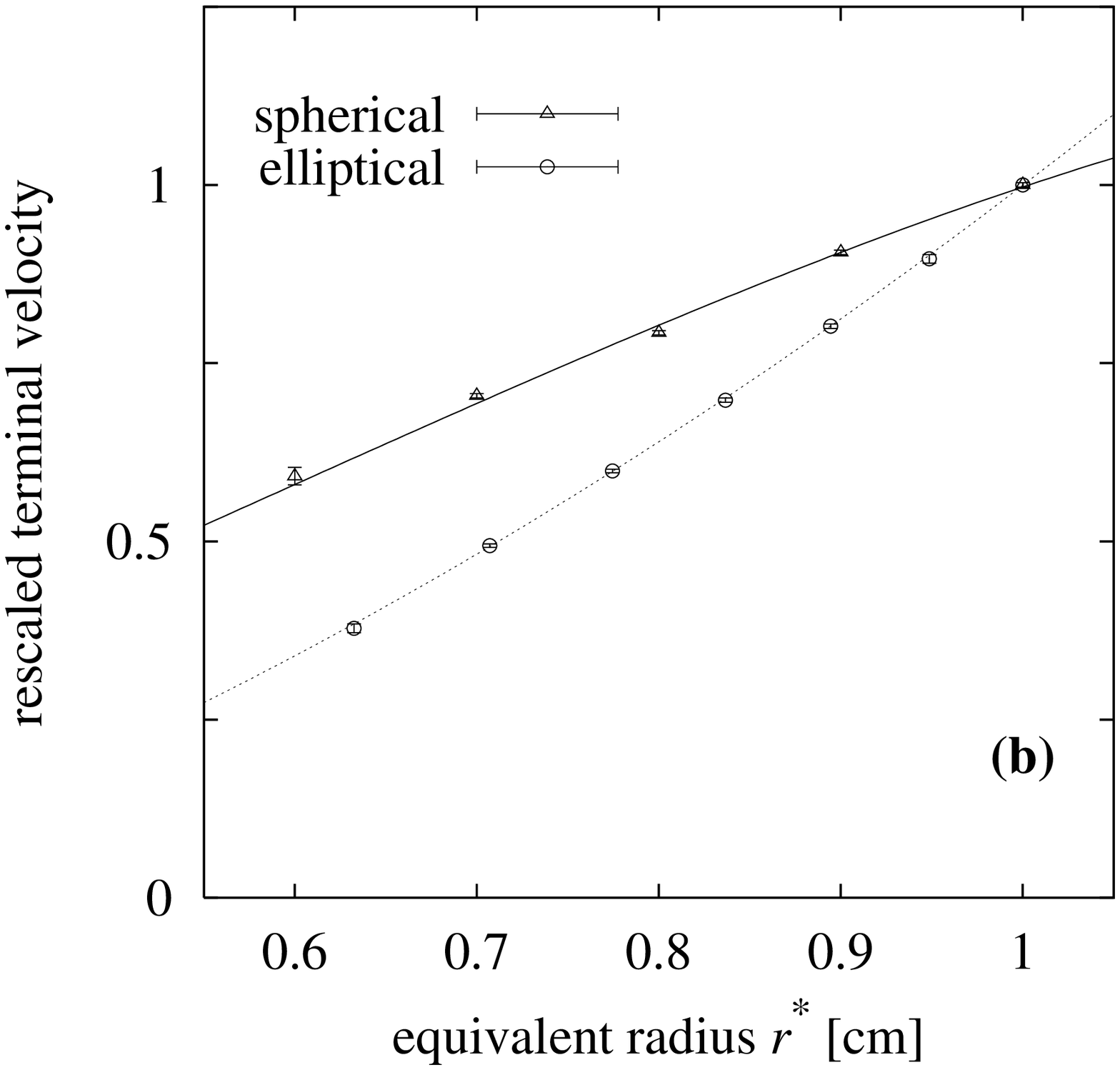} }
  \else
  \parbox[t]{0.495\textwidth}{
    \includegraphics[width=0.49\textwidth]{sink_ell1.pdf}}
    \hfill
  \parbox[t]{0.495\textwidth}{
    \includegraphics[width=0.49\textwidth]{sink_ell3.pdf} }
  \fi
  \fcaption{Terminal velocity of ellipses: (a) As function of the aspect ratio 
            and (b) as function of the corresponding equivalent cylinder with a
            spherical cross-section $d=2 r^*$ and for a system size of
            $L=15$\,cm. For comparison, we shown in (b) also the values for a
            spherical particle and two polynomial fits to guide the eye. 
            The error bars are roughly of the order of the symbol size.}
  \label{fig: vel_ell1}
\end{figure}

\noindent
For the simulation parameters used in this article, we also found that 
elliptical particles tend to align their minor axis with gravity in the long
run. In order to avoid long transition periods, we therefore start the
elliptical particle with an angle of the major axis with the horizontal of
0$^\circ$ to determine the terminal velocity for different aspect ratios AR =
$a/b$ where $a$ and $b$ denote the minor and major axis of the elliptical
cross-section, respectively. In Fig.~\ref{fig: vel_ell1}a, the dependence of the
terminal velocity on the aspect ratio is shown for three different system sizes
$L$. As in the case for a spherical particle, see Fig.~\ref{fig: validate}a, the
velocity increases with increasing system size since the drag force due to the
back-flowing fluid decreases.

The surface area of the cross-section of the elliptical particle is given by
\[  F_e = \pi a\, b \]
whereas the surface area of a spherical particle with radius $r$ is given by
\[  F_s = \pi r^2 \ . \]
In order to compare the dynamics of elliptical and spherical particles, one
defines a surface equivalent spherical particle for the ellipse with radius
\[ r^* = \sqrt{a\, b} \]
which has the same surface area as the elliptical particle.

This comparison is shown in Fig.~\ref{fig: vel_ell1}b for a system size of
$L=15$\,cm and a viscosity of $\eta=5$\,g/(cm s). The data points were
approximated by polynomial fits to guide the eye. The terminal velocity for
spherical cylinders is always {\em higher}\/ than the one for elliptical
particles when the surface equivalent spherical cylinder with radius $r^*$ is
considered. Due to the fact that the broad side of the ellipse is always turned
into the stream, the terminal velocity is nearly reduced by 50\% for a value of
$r^*=0.55$\,cm ($b=1$\,cm). The terminal velocities for the spherical and
elliptical cross-section shown in Fig.~\ref{fig: vel_ell1}b were rescaled by
the value for the unit spherical cross-section.

\subsection{Tumbling Motion}
\noindent
If an elliptical particle is released in a finite container filled with a
viscous fluid under gravity, the particle will for high enough Reynolds numbers
in general undergo a tumbling motion until its minor axis is perfectly aligned
with the direction of gravity and a translational motion until it sinks in the
middle of the container.\cite{feng94} The latter is given by the symmetry of
the problem and is also true for spherical cylinders.\cite{hu92} Numerical
simulations indicate that below a critical Reynolds number, an elliptical
particle will turn vertical in a Newtonian fluid.\cite{huang98} However, in our
numerical simulations we did not find such a behavior which might be due to the 
different simulation parameters.

\begin{figure}[t]
  \parbox[b]{0.48\textwidth}{
  \parbox{0.235\textwidth}{\centerline{AR=0.9}}
  \parbox{0.235\textwidth}{\centerline{AR=0.6}}\\[0.1cm]
  \ifx\pdfimage\undefined
    \includegraphics[width=0.235\textwidth]{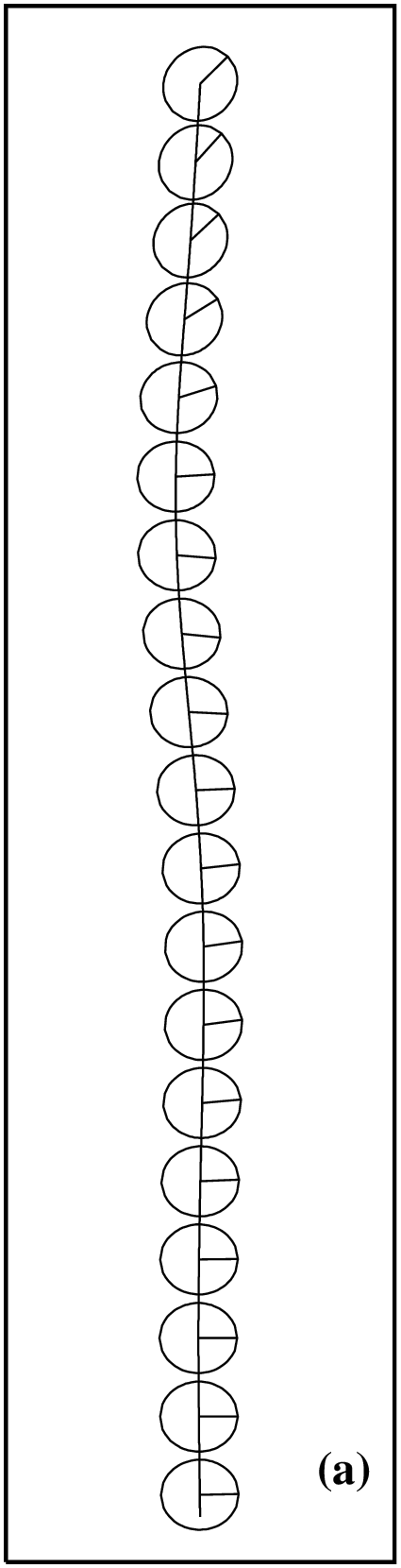}
    \includegraphics[width=0.235\textwidth]{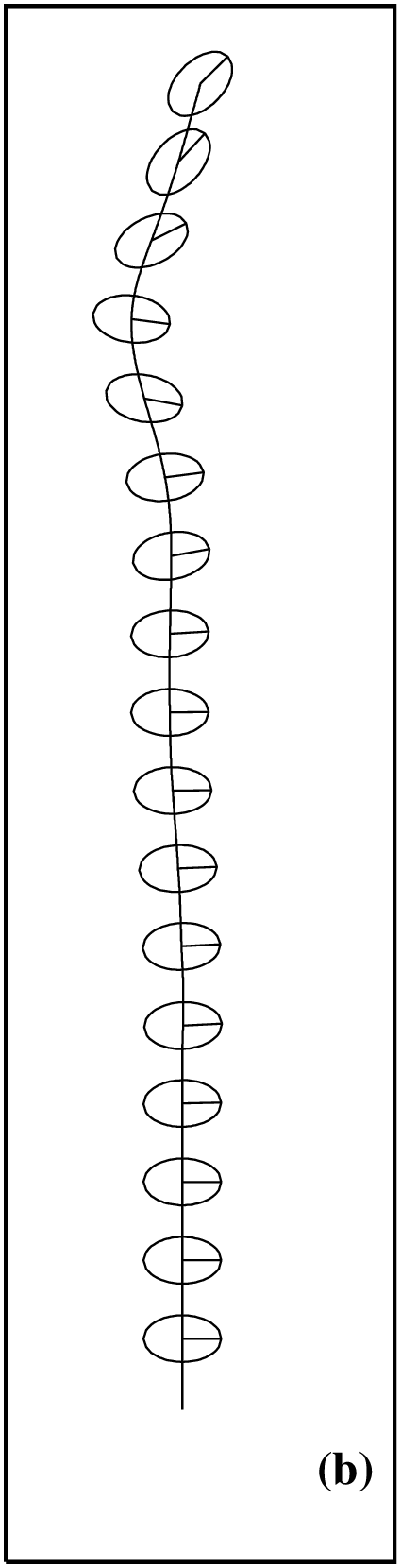}
  \else
    \includegraphics[width=0.235\textwidth]{rot_09_10cm.pdf}
    \includegraphics[width=0.235\textwidth]{rot_06_10cm.pdf}
  \fi }
  \parbox[b]{0.019\textwidth}{}
  \parbox[b]{0.5\textwidth}{
  \ifx\pdfimage\undefined
  \includegraphics[width=0.5\textwidth]{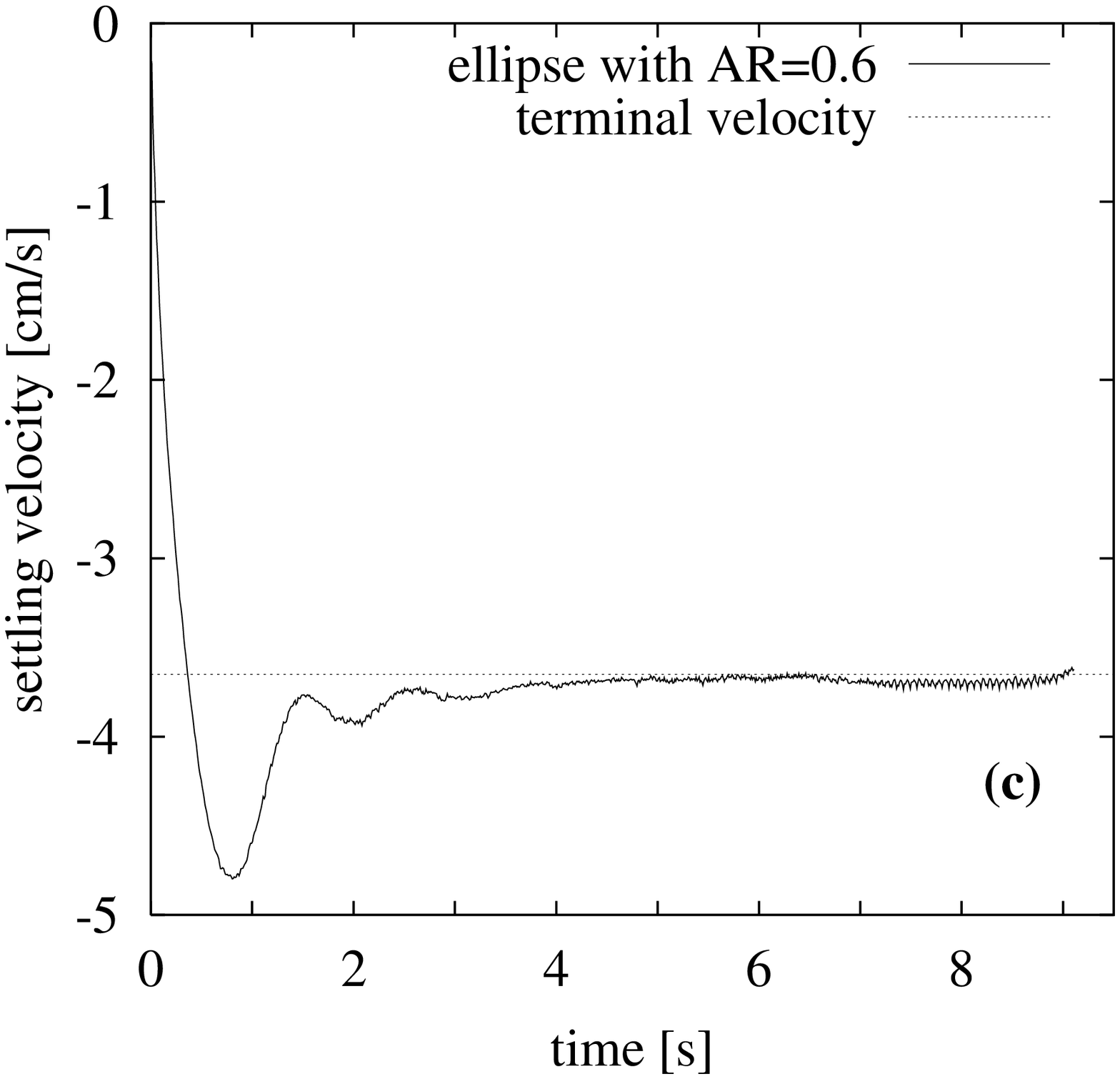}\\
  \includegraphics[width=0.5\textwidth]{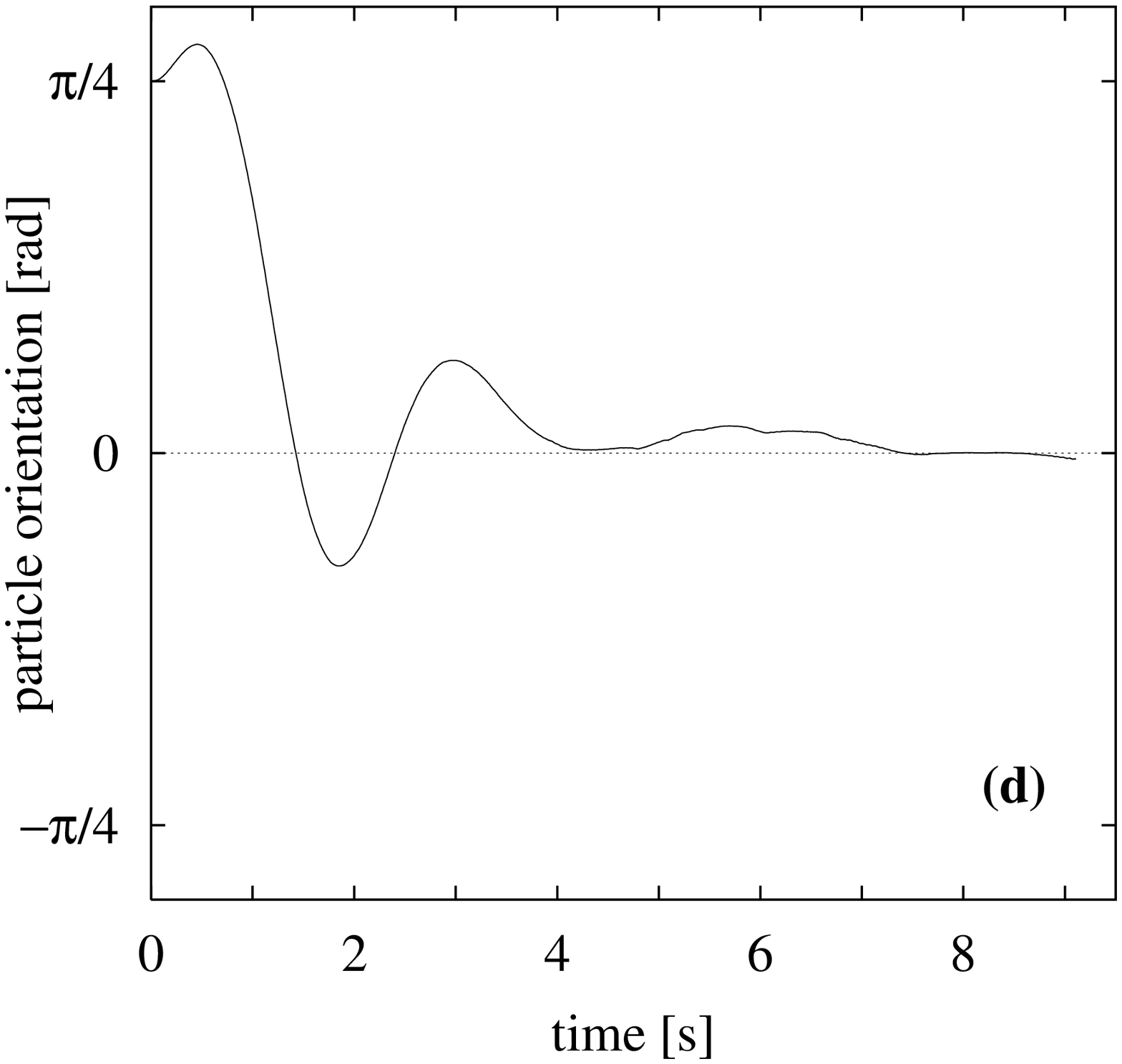}
  \else
  \includegraphics[width=0.5\textwidth]{rot_06_10cm_1.pdf}\\
  \includegraphics[width=0.5\textwidth]{rot_06_10cm_2.pdf}
  \fi }
  \fcaption{Settling dynamics of elliptical particles having an aspect ratio of
            0.9 (a) and 0.6 (b). For the latter case the settling velocity as
            function of time is shown in (c) and the particle orientation as
            function of time in (d). Please note the correlations between the
            orientation and the velocity.}
  \label{fig: vel_ell2}
\end{figure}

To illustrate this point, we show in Figs.~\ref{fig: vel_ell2}a,b the settling
motion of two elliptical particles in a container of size 10x40\,cm having an
aspect ratio of 0.9 and 0.6, respectively. Both particles were initially
released at an angle of $\pi/4$ (45$^\circ$) and the tumbling motion is clearly
visible in both cases but being more pronounced for lower aspect ratios (e.g.\
Fig.~\ref{fig: vel_ell2}b). Also note the translational motion which drags the
particle towards the centerline of the container.

For the particle motion depicted in Fig.~\ref{fig: vel_ell2}b, the settling and
tumbling dynamics are quantified by showing the settling velocity as function of
time in Fig.~\ref{fig: vel_ell2}c and the particle orientation, minor axis
with respect to gravity, as function of time in Fig.~\ref{fig: vel_ell2}d. The
latter graph shows how the oscillations in the orientation decrease in time and
how they overshoot the value of zero after around 2\,s after the particle was
released. This overshooting seems to be more pronounced and the decay seems to
be less pronounced when the aspect ratio is decreased from 0.9 to 0.7. However,
the opposite tendency is observed upon decreasing the aspect ratio even further
down to 0.6 which is due to the fact that the particle comes closer to the rigid
vertical side walls. It would be interesting to better quantify how the 
oscillations decay as function of aspect ratio and container width where
investigations are underway.

The settling velocity as function of time for an elliptical particle with an
aspect ratio of 0.6 is shown in Fig.~\ref{fig: vel_ell2}c. In the long run, the
velocity approaches the terminal velocity of $-3.65$\,cm/s which was added to
the figure as dashed line. Since the initial particle orientation was $\pi/4$,
the particle first moves to the left. During this motion, the angle decreases
and when it passes through zero, a force component to the right, towards the
center of the container, sets in. Please note how the particle orientation and
the settling velocity are related leading to velocity maxima shortly after the
particle orientation deviates the most from zero, i.e.\ comes the closest in
turning the thin side into the stream.

\section{Conclusions}
\noindent
We presented an efficient numerical algorithm to study the motion of spherical
and elliptical particles in two-dimensional viscous fluids. We validated our
algorithm by comparing its results with analytic expressions for fixed and
moving particles. in the latter case, good agreement with the Oseen
approximation is found in the low Reynolds number limit. However, our algorithm
is not restricted to this limit since it solves the full Navier-Stokes equations
and can easily be extended to deal with more than one particle.

This procedure was then used to study the settling and tumbling motion of
elliptical particles. The terminal velocity was compared to the one of the
surface equivalent sphere where we found that the terminal velocity is
drastically decreased due to the fact that the broad side of the ellipse is
always turned into the stream. It was also found that the magnitude and the
decay of the tumbling motion depends on the aspect ratio of the ellipse and the
container width.

\nonumsection{Acknowledgements}
\noindent
Stimulating and very helpful discussions with N. Lu are greatfully acknowledged.

\nonumsection{References}

\end{document}